\documentclass[10pt]{iopart}
\usepackage{epsf,citesort}

\newcommand{\gl}[1]{Eq.~(\ref{#1})}

\def\gtrless{\raise2.5pt\hbox{$>$}\llap{\lower2.5pt\hbox{$<$}}}
\def\gtrapprox{\raise2.5pt\hbox{$>$}\llap{\lower2.5pt\hbox{$\approx$}}}
\def\gtrapprox{\raise2.5pt\hbox{$>$}\llap{\lower2.5pt\hbox{$\approx$}}}
\def\lessapprox{\raise2.5pt\hbox{$<$}\llap{\lower2.5pt\hbox{$\approx$}}}
 
\newcommand{\beq}[1]{\begin{equation}\label{#1}}
\newcommand{\eeq}{\end{equation}}

\newcommand{\gd}{\dot{\gamma}}

\renewcommand{\rho}{\varrho}

\newcommand{\eps}{\varepsilon}

\begin{document}
\maketitle
\title[Non--Newtonian viscosity]{Non--Newtonian viscosity of
interacting Brownian particles: comparison of theory and data}

\author{Matthias Fuchs\dag and Michael E. Cates\ddag}

\address{\dag\ Institut Charles Sadron, 6, rue Boussingault, 67083
Strasbourg Cedex, France; permanent address: Physik-Department,
Technische Universit\"at M\"unchen, James-Franck-Str., 85747 Garching,
Germany}  

\address{\ddag\ Department of Physics and Astronomy, The University of
Edinburgh, JCMB King's Buildings, Edinburgh EH9 3JZ, GB}

\begin{abstract}
A recent first--principles approach to the non--linear rheology
of dense colloidal suspensions is evaluated and compared to simulation
results of sheared systems close to their glass transitions.
The predicted scenario of a universal transition of the structural
dynamics between yielding of glasses and  non--Newtonian (shear-thinning) fluid
flow appears well obeyed, and calculations within simplified models
rationalize the data over variations in shear rate and viscosity
of up to 3 decades.
\end{abstract}

\section{Introduction}

The rheological properties of soft materials, such as colloidal
dispersions, presumably originate in a number of physical mechanisms,
like shear--induced phase transitions, direct potential and
hydrodynamic interactions,  
advection of fluctuations, and shear banding or localization among others; see
e.~g. the collection of papers in \cite{Faraday123}. At higher
particle concentrations, the non--linear rheology depends on how
steady shearing interferes with solidification during glass formation.
Recently, we developed a theory for the non-linear rheology
of dense colloidal suspensions aimed at this point \cite{Fuchs02}. It
describes how the structural dynamics is fluidized by advection of density
fluctuations, while hydrodynamic interactions, non--linear flow profiles and
ordering phenomena are neglected.  Computer simulation studies 
can ensure that the latter processes are absent and 
thus provide crucial tests of the presented scenario. In this
contribution, theoretical calculations will be compared to Brownian dynamics
simulations of hard spheres by Strating \cite{Strating99} --- without
adjustable parameter in principle ---, and to molecular 
dynamics simulations of a sheared binary Lennard--Jones mixture by
Berthier and Barrat \cite{Berthier02}.

\section{Theory}
\label{theorie}

\subsection{General aspects}

A system of Brownian particles is studied in a  prescribed steady
shear solvent flow with constant velocity gradient and shear rate $\gd$. 
The equation of motion for the temporal evolution of the
many--particle distribution function  is known
\cite{dhont}, and has been solved for hard spherical particles at low
densities \cite{Bergenholtz01c}. 
This model constitutes a first microscopic approach to real
dense colloidal suspensions, and may serve as a model sheared glassy
fluid \cite{Fielding00}. It considers the ``Brownian part'' of the
viscosity only, which, in Stokesian Dynamics simulations,
  Foss and Brady found to dominate  
compared to the hydrodynamic one for small shear rates  \cite{Foss00}

While the (approximate) approach developed in \cite{Fuchs02}
gives general steady state
quantities (like the shear distorted static structure factor) 
and their time--dependent fluctuations close to glassy
arrest, we will concentrate on the thermodynamic shear
stress $\sigma(\gd)$ and the connected shear viscosity
$\eta(\gd)=\sigma/\gd+\eta_\infty$; here $\eta_\infty$ is the
viscosity of the background solvent. 
The equations of motion exhibit a glass transition bifurcation,
around which asymptotic expansions capture the transition from
shear--thinning fluid flow to solid--like yielding. With the
separation parameter $\eps$
denoting the (relative) distance from the transition, and $t_0$ a
time scale obtained by matching onto  microscopic transient motion,
the following behaviors 
of $\sigma$ in the ``structural window'' have been established \cite{Fuchs02}
\beq{e1}
\sigma= \sigma(\gd t_0,\eps) \to\left\{\begin{array}{ll}
\gd t_0 \; |\eps|^{-\gamma} \; c_1 & \eps < 0 \\
c_2 \; \left( 1 + c_3 |\gd t_0 |^m \right) & |\eps| \ll |\gd
t_0|^{\frac{2a}{1+a}} \\
c_2 \; ( 1 + c_4 \sqrt{\eps} ) &  \eps > |\gd
t_0|^{\frac{2a}{1+a}} \end{array}\right. \; ,
\eeq
where the $c_i$ are positive material--dependent  parameters (for the
exponents $\gamma$ \& $a$ see e.g. \cite{Franosch97}). 
The ``structural window'', here, is defined as the double regime
$|\eps| \ll 1$ and $|\gd t_0| \ll 1$, where the slowing--down of the
structural 
dynamics dominates the steady state stress. While the divergence of
the Newtonian viscosity $\eta_0=t_0 |\eps|^{-\gamma} c_1$ (first line
of \gl{e1}) upon approaching the transition, applies to the
linear--response regime of a fluid ($\eps<0$),  
and is known from mode coupling theory (see the references in
\cite{Franosch97}), the novel predictions close to and above
($\eps\ge0$) the transition describe the universal non--linear response
of glasses to steady shearing with rate $\gd$. Importantly, a ``dynamic
yield stress'' $\sigma^+(\eps)=\sigma(\gd\to0+,\eps\ge0)$ is obtained because
a finite stress has to be overcome in order to force the glass to
yield even for vanishingly small shear rate; $\sigma^+$ is connected
to the constants $c_2$ and $c_4$ in \gl{e1}.  While the yield stress
varies strongly with distance to the transition deep in the glass,
at fixed parameters close to the transition, the stress increases from
$\sigma^+$ with a  power--law in $\gd$,
 where the material--dependent exponent $m$ lies around
0.15 in the models studied below. The given asymptotes are only the
leading orders for $\eps\to0$ and $\gd t_0\to0$, while corrections
can be obtained systematically \cite{Fuchs02}, or are included in
model calculations to be presented below. 

The dominance of the structural dynamics in determining
 $\sigma(\gd t_0,\eps)$ entails
that all exponents or constants are functions of the equilibrium
structure factor $S_q$ alone, except for the time scale $t_0$ which matches
to shorter non--structural dynamics. Thus, hydrodynamic interactions
or inertial terms only influence the value of $t_0$, which ideally
could be determined from an analysis of the intermediate scattering
functions of the system \cite{Franosch97}.  This result arises because
the small--shear rate rheology of glassy suspensions is dominated by
steric hindrance (the ``cage--effect'') which is not qualitatively
affected by the properties of the solvent around the particles.
It is in agreement with the findings in Stokesian dynamics simulations
\cite{Foss00} that shear thinning is dominated by a decrease of the
Brownian part of the viscosity.  The elimination of particle
forces in favour of the quiescent--state structure factor $S_q$ is an
approximation of unknown quality in the present situation, but in part
motivated by the consideration of small shear rates.

\subsection{Models and simplifications}

The equations of motion, from which $\sigma$ in \gl{e1} follows
uniquely for a given $S_q$, have not been solved yet. 
Two approximate models were presented and discussed in
\cite{Fuchs02} and shall be used in the following. While the schematic
$F_{12}^{(\gd)}$--model only incorporates the competition of two effects
(divergent structural relaxation times with increasing $\eps$ and loss
of memory induced by shearing), the semi--microscopic ISHSM combines a
semi--quantitative description of a quiescent hard sphere colloidal
dispersion \cite{Franosch97} with an  isotropically--averaged shear advection
of density fluctuations. Both models only depend on two parameters
which map onto $\eps$ and $\gd$ introduced in \gl{e1}, and thus can be
viewed as minimal models for the described scenario. 

A problem when analysing data using both  models arises from the
ratio $c_2/c_1$ in \gl{e1}, which has a simple physical meaning. It
gives the ratio of yield stress to transverse 
elastic constant (viz. shear modulus $G_\infty$) of the glass at the
critical point, $c_2/c_1= \hat c_1 \sigma^+_c/G_\infty^c$, where the
numerical constant $\hat c_1=$ 1.0 (1) for the ISHSM
($F_{12}^{(\gd)}$). This ratio can be interpreted as a critical yield
strain.  Both models 
underestimate the effect of shearing leading to
$\sigma^+_c/G_\infty^c=$ 0.33 (0.34) for the ISHSM ($F_{12}^{(\gd)}$),
while experiments give values around 0.05 indicating smaller strains are
necessary for yielding \cite{russel}. While the schematic
$F_{12}^{(\gd)}$--model is not meant to quantitatively capture such
ratios, this error in the ISHSM presumably arises from the
oversimplified handling of the shear--induced anisotropy of density
fluctuations. The ISHSM treats all directions equivalent to the
vorticity direction that is perpendicular to the flow plane. Perhaps
unsurprisingly this underestimates the effects of shearing. We correct
for this error in an ad--hoc fashion by rescaling the shear--rate
$\gd$ when considering $\eta(\gd)$. For the $F_{12}^{(\gd)}$--model
this procedure is rigorously equivalent to an adjustment of the ratio
$\sigma^+_c/G_\infty^c$.

\section{Results and comparison with simulation data}

\begin{figure}
\begin{center}\epsfxsize=0.8\hsize
\epsfbox{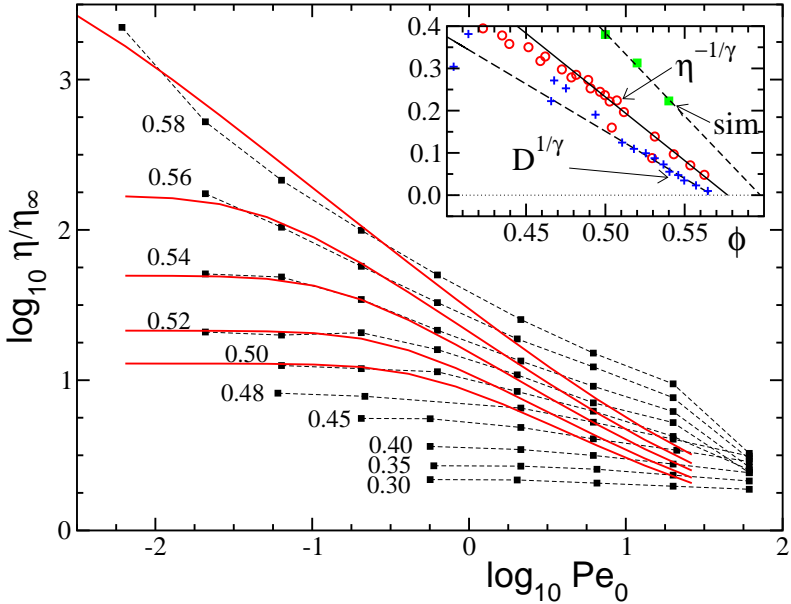}
\end{center}
\caption{\label{fig1} Steady state viscosities (symbols) from Brownian
dynamics simulations \protect\cite{Strating99} versus Peclet number
Pe$_0=\gd d^2/D_0$ for packing fractions $\phi$ as labeled.
Fits by eye to the data for $\phi\ge0.50$ with the
ISHSM for separation parameters $-\eps=$ 0.014, 0.058, 0.097, 0.139 \&
0.174 are given as solid lines and extrapolate to $\phi_c=0.59$. The
matching time  $t_0=0.019 d^3\eta_\infty/k_BT$ is obtained and the
theoretical $\gd$ is 
rescaled to $0.25 \gd$ as discussed in the text. The inset shows a
rectification plot with predetermined exponent $\gamma=$ 2.62 
of viscosities
from experiments \protect\cite{Cheng02} (circles) and simulations
\protect\cite{Strating99} (squares), alongside self diffusion constants
(crosses) from \protect\cite{Megen98}, versus
packing fraction. Linear fits to the data above $\phi\ge0.50$
give  $(\eta/\eta_\infty)^{-1/\gamma}=1.2\, \eps$ with $\phi_c=$ 0.58
(two outliers neglected), $(\eta/\eta_\infty)^{-1/\gamma}=1.6 \,\eps$
with $\phi_c=$ 0.60, and $(D_L/D_0)^{1/\gamma}=0.8 \, \eps$ with $\phi_c=$ 0.57
respectively.}
\end{figure}

Before applying \gl{e1} to a solution of colloidal hard spheres at
packing fraction $\phi$, at
first the latter's critical value, $\phi_c$, entering in $\eps=C
(\phi_c-\phi)/\phi_c$  (with $C=1.5$ \cite{Franosch97}) needs to be
determined. This is done by testing whether 
the divergence of the quiescent viscosity (and corresponding
structural relaxation time) for
$\eps\to0-$ is observed. The inset of fig. \ref{fig1} shows
viscosities from experiments  \cite{Cheng02} and from Brownian
dynamics simulations \cite{Strating99}. Also included are 
self--diffusion coefficients from  \cite{Megen98}, which are predicted
to vanish with $D\propto|\eps|^\gamma$. Replotting the data with the calculated
$\gamma=2.62$ \cite{Franosch97}, fits to the data above
$\phi\ge0.50$ give $\phi_c=$ 0.57 \cite{Megen98}, 0.58 \& 0.60 for
$D$, $\eta$ (two outliers 
neglected) and simulations, respectively.  Interestingly, the two
experimental data sets provide rather close estimates for $\phi_c$ and
indicate a strong coupling of diffusion and viscosity, $D\eta \to
0.4 \eta_\infty D_0$ for $\eps\to0$ (neglecting the difference in
$\phi_c$), with $D_0$ the dilute single particle diffusion 
coefficient. The numerical factors is about half the predicted value
\cite{Franosch97}. We speculate that the discrepancy of the
extrapolation of the simulation results
arises in part because the data are not fully in the asymptotic
regime.

With the quantitative knowledge of $\eps$, only the matching time
$t_0$ is required to analyse the steady state viscosities in the
structural window using the ISHSM. We chose to obtain it via the full
fitting procedure which consists in matching by eye the
numerical solutions to the non--Newtonian viscosity data.  
In this way, $t_0$ is mainly determined by the increase of the Newtonian
viscosity, because  $\eta\sim t_0\; \sigma/(\gd t_0)$ holds
and $\sigma/(\gd t_0)$ becomes independent of $t_0$ in the fluid for
vanishing shear rate. The main panel of fig. \ref{fig1} shows
$\eta$ from the Brownian dynamics simulations as function of the
dimensionless Peclet 
number Pe$_0=\gd d^2/D_0$, which measures the effect of shearing
relative to the time a single particle diffuses its diameter $d$.
The fits by eye using the ISHSM are included for packing fractions close 
to the transition,  $\phi\ge0.50$. From the fits, mainly from the
divergence of $\eta_0$ given in \gl{e1},  the
matching time $t_0= 0.019 d^3\eta_\infty/k_BT$ is estimated, and inclusion of
corrections to asymptotic behavior in the ISHSM--fits shifts the glass
transition packing fraction closer to the other determinations;
$\phi_c=0.59$ follows from the  $\eps$ used in fig. \ref{fig1}.
Note that the solvent viscosity is included in the theoretical curves,
$\eta=\eta_\infty+ t_0\, \sigma/(\gd t_0)$\footnote{The ISHSM calculations
provide $\sigma(\eps,\gd t_0,$Pe$_0)$ for all values of $\eps$ and
Pe$_0$, while \gl{e1} captures the asymptotic behavior for $\eps\to0$,
$\gd t_0\to0$ \& Pe$_0\to0$. Because we aim at describing the
proximity of the glass transition, we match the parameters of \gl{e1}
($\phi_c$ and $t_0$) there. Without matching the ISHSM gives
$\phi_c=0.52$ \& $t_0=0.025 d^3\eta_\infty/k_BT$. }.  
In the shear--thinning region, the viscosity diminuishes and
approaches a behavior like $\eta\sim \sigma_c^+/ \gd$ with strong
corrections, though, masking the power--law \cite{Fuchs02}. Because of
the overestimate of $\sigma_c^+$ in the ISHSM, this decrease would set in at
too high $\gd$ values only. In order to correct for the quantitative
error, the theoretical curves are plotted versus rescaled shear rate,
$\gd*0.25$; i.e. $\sigma/(\gd t_0)=f_\eta(\gd*0.25)$. 
With this ad--hoc correction, satisfactory agreement of theory and
simulation results is seen for Pe$_0\le 1$, where the steady state
viscosity varies over two orders on variation of shear rate and
packing fraction.  
For larger Peclet numbers, the data presumably lies outside 
the structural window where \gl{e1} applies.
Motivated by numerical findings in \cite{Fuchs02},
we speculate that the enhanced, $\phi$--dependent steady state
viscosities around Pe$_0=$ 1 -- 10  in 
the Brownian dynamics simulations originate in the hard core
repulsion. If so, hydrodynamic interactions which prevent particles
from close contact could appreciably affect $\eta$ in this region.

\begin{figure}
\begin{center}\epsfxsize=0.8\hsize
\epsfbox{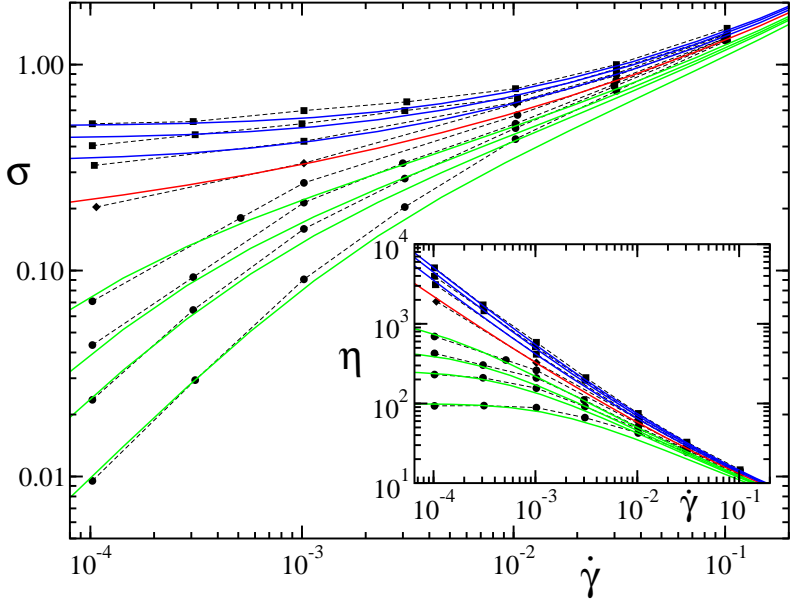}
\end{center}
\caption{\label{fig2}Symbols are shear stress (main panel) and viscosity
(inset) data of a super--cooled Lennard--Jones binary mixture in
reduced units taken from Ref. \cite{Berthier02}; from top to bottom,
the temperatures are  0.15, 0.3, 0.4, 0.45, 0.5, 0.525, 0.555 \& 0.6
while $T_c\approx 0.435$. The solid lines give fits by eye using the 
F$_{12}^{(\gd)}$--model from Ref. \cite{Fuchs02} with separation
parameters: $\eps=$ 0.050, 0.037, 0.021, 0.0, $-0.027$, $-0.042$,
$ -0.054$ \& $-0.083$
    (from top to bottom); units are converted by $\sigma=\gd\eta=1.8 \langle
\tau \rangle \gd$, where $\gd=$ 0.53 Pe$_0$, and $\langle \tau\rangle=
\langle \tau(\eps,$Pe$_0)\rangle$ \cite{Fuchs02}.}
\end{figure}

A second set of steady state shear stresses and viscosities is
provided by recent large scale molecular dynamics simulations of a
sheared simple liquid (a binary Lennard--Jones mixture) above and 
below its glass transition temperature \cite{Berthier02}. Because
kinetic parameters do not enter the theoretical predictions, and as
linear flow profiles were obtained in the simulations, the
universal predictions of our approach can be compared again. Figure
\ref{fig2} shows stationary shear stresses from the simulations and
fits by eye using the $F_{12}^{(\gd)}$--model as specified in
\cite{Fuchs02}. The model provides a relaxation time $\langle
\tau\rangle$ as function of $\eps$ and a dimensionless shear rate,
denoted as Peclet number Pe$_0$, which are mapped onto the data as
specified in the figure caption.  The data nicely span the glass
transition temperature, $T_c\approx 0.435$ already known
\cite{Berthier02}, and are well compatible with a transition from a
shear--thinning fluid to a yielding glass with finite yield stresses
at and below the transition temperature.

\section{Conclusions and outlook}

We presented results of a microscopic theory of the
nonlinear rheology of colloidal fluids and glasses under steady
shear \cite{Fuchs02}, and compared them with simulation and experimental data.
This brought out the existence of a
universal transition between shear-thinning
fluid flow, with diverging viscosity upon increasing the interactions,
and solid yielding, with a yield stress that is finite at (and beyond)
the glass point.  Numerical calculations could explain simulation
results over up to 3 decades in variation in shear rate and viscosity.
A quantitative analysis of larger data sets is required in order to
determine the theoretical parameters for both simulations more
accurately than the estimates found here.

The approach we outlined should be improved with respect to the handling of
shear--induced anisotropies, and stress--induced effects. The latter
may lead to 
shear thickening behaviour that, for many colloidal materials, occurs
at higher flow rates than those addressed here. This avenue will be
explored in a future paper \cite{Holmes02} on a version of the schematic
model which is modified to include explicit stress- (as well as
strain--rate--) dependence.

\ack 
We thank J.--L. Barrat and L. Berthier for discussions.
M.F.\ was supported by the DFG, grant Fu~309/3. 

\vspace{\baselineskip}
 
\hrule
 
\vspace{\baselineskip}


\begin{thebibliography}{10}

\bibitem{Faraday123}
Pusey P N, ed
\newblock  2002 {\em Non-Equilibrium Behaviour of Colloidal Dispersions},
  Faraday Disc. {\bf 123}

\bibitem{Fuchs02}
Fuchs M and Cates M~E
\newblock  2002 {\em cond-mat/}0204628; {\em Faraday Disc.} {\bf 123}
in print \& {\em cond-mat/}0207530   

\bibitem{Strating99}
Strating P
\newblock  1999 {\em Phys. Rev. E} {\bf 59} 2175

\bibitem{Berthier02}
Berthier L and Barrat J~L
\newblock  2002 {\em J. Chem. Phys.} {\bf 116} 6228

\bibitem{dhont}
Dhont J K~G
\newblock  1996 {\em An introduction to dynamics of colloids}
\newblock (Elsevier Science, Amsterdam)

\bibitem{Bergenholtz01c}
Bergenholtz J, Brady J~F  and Vicic M
\newblock  2002 {\em J. Fluid Mech} {\bf 456} 239

\bibitem{Fielding00}
Fielding S~M, Sollich P  and Cates M~E
\newblock  2000 {\em J. Rheology} {\bf 44} 323

\bibitem{Foss00}
Foss D and Brady J F \newblock  2000 J. Fluid Mech {\bf 407} 167 

\bibitem{Franosch97}
Franosch T, Fuchs M, G{\"o}tze W, Mayr M~R  and Singh A~P
\newblock  1997 {\em Phys. Rev. E} {\bf 55} 7153; and references therein

\bibitem{russel}
Russel W~B, Saville D~A  and Schowalter W~R
\newblock  1989 {\em Colloidal Dispersions}
\newblock (Cambridge University Press, New York)

\bibitem{Cheng02}
Cheng Z, Zhu J, Chaikin P~M, Phan S-E  and Russel W~B
\newblock  2002 {\em Phys. Rev. E} {\bf 65} 041405

\bibitem{Megen98}
van Megen W, Mortensen T~C, Williams S~R  and M{\"u}ller J
\newblock  1998 {\em Phys. Rev. E} {\bf 58} 6073

\bibitem{Holmes02}
Holmes C~B, Fuchs M  and Cates M~E
\newblock  2002 {\bf in preparation}

\end{thebibliography}

\end{document}